# Interlayer Interactions in Anisotropic Atomically-thin Rhenium Diselenide

Huan Zhao[1,ł], Jiangbin Wu[2,ł], Hongxia Zhong[3,5], Qiushi Guo[4], Xiaomu Wang[4], Fengnian Xia[4], Li Yang[3], Ping-Heng Tan[2,*], Han Wang[1,*]

[1]*Ming Hsieh Department of Electrical Engineering, University of Southern California, Los Angeles, CA 90089, USA*

[2]*State Key Laboratory of Superlattices and Microstructures, Institute of Semiconductors, Chinese Academy of Sciences, Beijing 100083, China*

[3]*Department of Physics, Washington University in St Louis, St Louis, MO 63130, USA*

[4]*Department of Electrical Engineering, Yale University, New Haven, CT 06511, USA*

[5]*State Key Laboratory for Mesoscopic Physics and Department of Physics, Peking University, Beijing 100871, China.*

**Abstract.** Recently, two-dimensional (2D) materials with strong in-plane anisotropic properties such as black phosphorus have demonstrated great potential for developing new devices that can take advantage of its reduced lattice symmetry with potential applications in electronics, optoelectronics and thermoelectrics. However, the selection of 2D material with strong in-plane anisotropy has so far been very limited and only sporadic studies have been devoted to transition metal dichalcogenides (TMDC) materials with reduced lattice symmetry, which is yet to convey the full picture of their optical and phonon properties, and the anisotropy in their interlayer interactions. Here, we study the anisotropic interlayer interactions in an important TMDC 2D material with reduced in-plane symmetry - atomically thin rhenium diselenide ($ReSe_2$) - by investigating its ultralow frequency interlayer phonon vibration modes, the layer-number dependent optical bandgap, and the anisotropic photoluminescence (PL) spectra for the first time. The ultralow frequency interlayer Raman spectra combined with the first study of polarization-resolved high frequency Raman spectra in mono- and bi-layer $ReSe_2$ allows deterministic identification of its layer number and crystal orientation. PL measurements show anisotropic optical emission intensity with bandgap increasing from 1.26 eV in the bulk to 1.32 eV in monolayer, consistent with the theoretical results based on first-principle calculations. The study of the layer-number dependence of the Raman modes and the PL spectra reveals the relatively weak van der Waal's interaction and 2D quantum confinement in atomically-thin $ReSe_2$. The experimental observation of the intriguing anisotropic interlayer interaction and tunable optical transition in mono- and multi-layer $ReSe_2$ establishes the foundation for further exploration of this material to develop anisotropic optoelectronic devices in the near-infrared spectrum range where many applications in optical communications and infrared sensing exist.

*Corresponding authors: H. W. (han.wang.4@usc.edu), P.-H. T. (phtan@semi.ac.cn)

ł These authors contributed equally to this work.

## Introduction

In the past few years, the transition metal dichalcogenides (TMDC) family[1-6] of two-dimensional (2D) materials have attracted great interest among the physicists, chemists, engineers and material scientists due to their unique physical and chemical properties resulting from 2D quantum confinements, unique lattice structures and interlayer coupling (or the absence of it in their monolayer form). Recently, 2D materials with strong in-plane anisotropic properties such as black phosphorus[7-11] have been proposed for developing new devices with promising applications in electronics,[12,13] optoelectronics[14-18] and thermoelectrics.[19] However, the selection of 2D materials with strong in-plane anisotropy has so far been very limited and only sporadic studies have been devoted to transition metal dichalcogenides materials with strong in-plane anisotropy, which is yet to convey the full picture of their anisotropic optical and phonon properties, and the anisotropy in their interlayer interactions.

In this work, we study the anisotropic interlayer interactions in an important TMDC 2D material with reduced symmetry – atomically-thin rhenium diselenide ($ReSe_2$) – by investigating its ultralow frequency interlayer phonon vibration modes, the layer-number dependent optical bandgap, and the anisotropic photoluminescence (PL) spectra for the first time. The ultralow frequency interlayer Raman spectra combined with the high-frequency Raman measurements allow deterministic identification of the $ReSe_2$ layer number and crystal orientation. The PL measurements show anisotropic optical emission intensity, which depends on the polarization of the incoming light with bandgap increasing from 1.26 eV in bulk to 1.32 eV in monolayer, consistent with theoretical results based on first-principle calculations using density functional theory (DFT). A systematic study of the polarization-resolved high frequency Raman spectra in mono- and bi-layer $ReSe_2$ is also carried out for the first time. Study of the layer-number dependence in the high frequency Raman modes and PL indicates relatively weak van der Waal's (vdW) interaction and 2D quantum confinement in atomically-thin $ReSe_2$. The results reveal the intriguing interlayer interaction and anisotropic optical transition in single- and multi-layer $ReSe_2$, which offers potential for developing anisotropic optoelectronic devices in the near infrared spectrum range.

$ReSe_2$ is one of the layered transition metal dichalcogenides (TMDCs) with van der Waals interaction between layers. As shown in Fig. 1a, every unit cell of monolayer $ReSe_2$ contains four formula units, which includes two categories of rhenium (Re) atoms together with four categories of selenium (Se) atoms. The Se atoms on top and bottom sandwich the Re atoms in the middle to form a monolayer lattice of $ReSe_2$. Unlike common TMDCs such as $MoS_2$ and $WSe_2$, which crystallize in the hexagonal (H) phases, $ReSe_2$ crystal displays a distorted $C_dCl_2$-type lattice structure.[20] Due to Peierls distortion,[21] adjacent Re atoms are bonded in the form of zigzag four-atom clusters,[22] which align along the direction of the lattice vector **a** to form Re chains (see Fig. 1b). Calculations have revealed that such a distorted octahedral (1T) crystal structure has lower energy than its hexagonal counterpart, thus being more stable.[23] This clustering of Re atoms has also been discovered in $ReS_2$ crystals,[20, 24] contributing to distort the lattice geometry and stabilize the crystals. The distorted 1T nature of $ReSe_2$ has conferred this materials strong in-plane anisotropy in their optical[25-27] and electronic[28-32] properties. Figure 1c shows the Brillouin zone of the monolayer $ReSe_2$, which is a hexagon with unequal side length. The band structure of monolayer $ReSe_2$ predicted by first-principle calculations is also demonstrated in Fig 1c. Perdew-Burke-Ernzerhof (PBE) exchange-correlation function under general gradient approximations (GGA) was applied in the calculations. Spin-orbit coupling (SOC) was taken into account and ultrasoft pseudopotential was applied. The calculated bandgap is 1.15 eV. It is well known that GGA tends to underestimate the bandgap of 2D materials,[33] so the real intrinsic bandgap is likely to be larger than 1.15 eV as confirmed by our experimental results to be discussed later. From the energy band diagram, the bottom of the conduction band is located at the Γ point, while the top of the valence band is located near the Γ point, making monolayer $ReSe_2$ an indirect bandgap semiconductor.

## Experiments and results:

To investigate the interlayer interactions in this 2D material with reduced symmetry, $ReSe_2$ flakes were prepared on $Si/SiO_2$ substrates by the standard micromechanical exfoliation method.[34] Mono- to few-layer samples were first located by optical contrast using optical microscope and then the layer numbers were verified based on the height information measured by Atomic Force Microscopy (AFM). Fig. 1d shows the optical micrograph and the AFM data for mono- and few-layer $ReSe_2$ flakes. According to the AFM morphology, the thickness of mono-layer $ReSe_2$ is about 0.7 nm, which is in agreement with the interlayer distance obtained by powder diffraction.[35] $ReSe_2$ samples in their mono- to few-layer forms are robust in air and show no clear signs of degradation after at least several months in ambient environment.

The interlayer shear (C) and layer-breathing (LB) Raman modes of a layered material directly reflect its interlayer van der Waals (vdW) coupling.[36-42] These modes only occur in multi-layers of $ReSe_2$, but not in the monolayer samples. As shown in Fig. 2a, the C and LB modes often occur at the ultralow frequency region (<50 $cm^{-1}$) of Raman spectra in TDMCs because the vdW coupling is a much weaker interaction compared to the intra-layer modes where the lattice atoms interact through the much stronger atomic bonding. In general, there are $N$-1 pairs of C modes and $N$-1 LB modes in an $N$ layer 2D layered material. For in-plane isotropic materials, like graphene[36] and $MoS_2$,[37] each pair of C modes is doubly degenerate and has the same frequency. For an in-plane anisotropic 2D material of $ReSe_2$, each pair of C modes is not degenerate in principle, which means that there are 2($N$-1) C modes with different frequencies in an $N$ layer sample and these 2($N$-1) C modes can be divided into two categories based on the vibration directions. Although $ReSe_2$ is an anisotropic material, the lattice constants of directions **a** (6.7 Å) and **b** (6.6 Å) are almost equal, which suggests that each pair of C modes of the material tend to overlap together because of their small frequency separation. Indeed, first principle calculations using DFT (Figure 2b) predict two C modes at 12.1 $cm^{-1}$ and 13.2 $cm^{-1}$, respectively, and a LB mode at 24.8 $cm^{-1}$. Figure 2a exhibits the ultralow frequency modes in 1L, 2L, 3L, 6L and 10L $ReSe_2$. The spectrum for bilayer $ReSe_2$ has the C mode at 13.5 $cm^{-1}$ and the breathing mode at 25.0 $cm^{-1}$. Since the theoretical difference between the two C modes in frequency are as small as 1 $cm^{-1}$, the peak measured at 13.5 $cm^{-1}$ for bilayer $ReSe_2$ should correspond to the overlap of the two calculated C modes at 12.1 $cm^{-1}$ and 13.2 $cm^{-1}$. Figure 2b shows the lattice vibration displacement directions for the three modes. The C and LB modes predicted by DFT calculations for trilayer $ReSe_2$ are included in the supplementary information (Fig. S3).

Given an $N$ layer $ReSe_2$, we use $C_{NN-i}$ and $LB_{NN-i}$ ($i$=1, 2…, $N$-1) to denote the $N$-1 C modes and $N$-1 LB modes. Here, $C_{N1}$ and $LB_{N1}$ (i.e., $i$=$N$-1) are the highest frequency C and LB modes, respectively. As shown in Figure 2a, no peaks corresponding to interlayer Raman modes were detected in the monolayer sample. The $C_{21}$ and $LB_{21}$ are observed at 13.5 $cm^{-1}$ and 25.0 $cm^{-1}$ in the bilayer flake, respectively. However, only $C_{32}$ (9.5 $cm^{-1}$) and $LB_{32}$ (17.7 $cm^{-1}$) are detected in trilayer $ReSe_2$. The $C_{31}$ and $LB_{31}$ modes are silent. Moreover, only the $C_{NN-1}$ and $LB_{NN-1}$ are observed in the 6 layer and 10 layer samples. The likely reason for the absence of other C and LB modes is their Raman-inactive nature or their much weaker electron phonon coupling compared to the lowest frequency modes, which is similar to the case of the highest frequency C modes in multi-layer graphene.[36]

Figure 2c shows the polarization dependence of the C and LB modes measured in bilayer and tri-layer $ReSe_2$. Due to the reduced in-plane symmetry in $ReSe_2$ lattice, the intensity of the C and LB modes show strong dependence on the polarization angle of the incident laser beam. This clearly reveals the interlayer coupling characteristics and the anisotropic interlayer vibration modes in $ReSe_2$, which suggests that the ultralow frequency interlayer Raman measurements can serve as an unequivocal way to determine the layer number in $ReSe_2$ samples and their crystal orientation.

The intra-layer Raman modes of layered materials correspond to the Raman-active intra-layer phonon vibration. The layer-number dependence of these high frequency Raman modes (>50 $cm^{-1}$) can also provide insight into the interlayer interaction in 2D materials. At the high frequency region of the monolayer $ReSe_2$ Raman spectra, in principle, there should be 36 normal vibration modes including 18 potential Raman modes due to the presence of 12 atoms in each unit cell of the $ReSe_2$ crystal lattice. We detected more than 10 distinctive Raman peaks in the spectral range between 100 $cm^{-1}$ and 300 $cm^{-1}$ in the Raman spectrum of monolayer $ReSe_2$, as depicted in Figure 3a.

Due to the reduced in-plane symmetry, the intensity of the intra-layer Raman modes in $ReSe_2$ also exhibits strong polarization dependence. By varying the polarization direction of the incident laser beam, we investigated the polarization dependence of various Raman modes of monolayer $ReSe_2$. Measurements were carried out from 0 degree to 360 degree with twenty-degree steps. Fig. 3a is the plot of Raman spectra under different polarization directions. There are no obvious shifts of peak positions when tuning the polarization direction of the incident laser beam. However, the peak intensity of all the Raman modes varies significantly, with a period of 180 degrees. This dependence can be clearly observed in the polar plots of the peak intensity as a function of the polarization angle. Fig. 3b shows the plots for the peaks at 125 $cm^{-1}$, 160 $cm^{-1}$, and 176 $cm^{-1}$. Similar plots of additional characteristic peaks are available in the supplemental materials (Fig. S2). Since modifying the relative angle between laser polarization direction and crystal orientation leads to variation of Raman intensities, we can utilize polarization-resolved Raman measurement as a non-destructive tool to identify the crystal orientation of $ReSe_2$. This is the first systematic study of high frequency Raman spectra in mono- and bi-layer $ReSe_2$. Our results are also consistent with a previous study[43] that showed Raman measurement data for 5 layer and thicker $ReSe_2$ samples.

To understand the corresponding lattice vibrations of each Raman mode, we calculated the phonon modes of monolayer $ReSe_2$. Among 36 normal modes of phonon vibrations, there are 18 Raman modes, 15 infrared modes, and three acoustic modes. Hence, the lattice vibration in each Raman mode can be rather complex. Density functional theories were applied to identify the vibrations of each Raman mode. The calculated modes match well with the measured peaks. A table (Table S2) comparing calculated and experimental peaks are provided in the supplementary information. We found that the vibration in each mode contains multiple components along different lattice directions. In addition, since there are four Se basis atoms and two Re basis atoms in each unit cell, the intensity of vibration could vary significantly in atoms with different positions. We selected the 125 $cm^{-1}$, 160 $cm^{-1}$, and 176 $cm^{-1}$ peaks and showed the lattice vibration and atom displacement directions for their corresponding phonon modes in Fig. 3c. Due to the complexity of lattice vibrations in monolayer $ReSe_2$, it is hard to find a pure $E_g$ or $A_g$ mode corresponding to each Raman peak as the case in $MoS_2$, thus we identify these modes as $E_g$-like or $A_g$-like modes based on their dominant vibration directions. The 125 $cm^{-1}$ mode is an $E_g$-like mode, as the vibration is mostly in-plane and symmetric. The 160 $cm^{-1}$ and 176 $cm^{-1}$ peaks are $A_g$-like modes since the main vibrations are in the one-dimensional vertical direction.

Figure 4 shows the layer-number dependence of the $ReSe_2$ Raman spectra for monolayer, bilayer and four-layer samples. Bulk $ReSe_2$ displayed very weak Raman intensity, compared to its few-layer counterparts. For all of the peaks, except the one near 176 $cm^{-1}$, Raman intensities of monolayer samples are slightly weaker than those in the 3-4 layer samples. This is attributed to its small volume in the unit cell and the multiple reflection interference in the multilayer structures contained $ReSe_2$ flake, $SiO_2$ and Si substrate, similar to the case of graphene multilayers.[44] For most of the Raman modes, thick samples have red-shifted peak positions compared to thinner samples. A detailed discussion of layer-number dependence of high frequency Raman modes is included in the discussions section.

The interlayer coupling and 2D quantum confinement in atomically thin $ReSe_2$ samples also result in the dependence of their bandgap on layer number, which can be revealed from photoluminescence (PL) measurements. We carried out PL measurements on monolayer, few-layer, and bulk $ReSe_2$ samples (Fig. 5a). For each sample, only one main PL peak was found. However, the PL spectrum of bulk $ReSe_2$ has a relatively broader peak, which is likely the combination of two split peaks. A detailed discussion of the PL split at low temperature can be found in the supplemental materials. Due to the indirect bandgap nature, the peak intensity is relatively weak for monolayer $ReSe_2$, and increases monotonically when adding layer numbers (Figures. 5b). Thus, monolayer samples are relatively noisy, but better signal-to-noise was obtained in few-layer samples. The measured bandgap enlarges with deceasing layer numbers, ranging from 1.26 eV of bulk to 1.32 eV of monolayer crystals (Fig. 5a, 5b), demonstrating the same trend in the layer-number dependence of the bandgap as predicted by previous first-principle calculations of $ReSe_2$ bandgap at different thicknesses.[23]

To further study the anisotropic optical absorption, we conducted the polarization-dependent PL measurements on few-layer $ReSe_2$ flakes by rotating the polarization of the incident laser beam. Since the PL signal of very thin $ReSe_2$ is relatively weak due to its indirect bandgap, we chose multi-layer samples to obtain higher signal-to-noise ratio. Multi-layer $ReSe_2$ regions of different thickness (6-layer and 10-layer) were selected from a single continuous flake sample, and thus should have the same crystal orientations, which is also confirmed by Raman measurements. By varying the polarization direction of the excitation laser beam, we were able to obtain the polarization dependent PL peak intensities. As shown in Fig. 5c, the peak intensity of two flake regions with 6-layer and 10-layer $ReSe_2$ vary with polarization angles at a period of 180°, revealing their anisotropic energy-band information.

## Discussions

**Ultralow-frequency Interlayer Raman modes.** Our ultralow frequency Raman measurement is the first experimental study of interlayer vibrations of few-layer $ReSe_2$. These interlayer Raman modes are important since they offer information about interlayer charge exchanges, screenings and scatterings. The frequency of the observable C and LB modes of multi-layer layered materials can be calculated by these two equations,[37]

$$\omega(C_{NN-i}) = \sqrt{2}\omega(C_{21})\sin(i\pi/2N) \quad \text{and}$$

$$\omega(LB_{NN-i}) = \sqrt{2}\omega(LB_{21})\sin(i\pi/2N)$$

respectively, where $N$ is the layer number, $i = 1,2,\ldots,N-1$. Using this formula, the layer number can be determined by the peak positions of the C and LB modes. Compared to AFM data, measurements of ultralow frequency Raman modes provide a more accurate method for identifying $ReSe_2$ layer numbers. Our first-principle calculations were able to accurately predict the C and LB Raman modes of bi-layer and tri-layer $ReSe_2$. Additional information about the calculated modes can be found in the supplemental materials (Table S1).

**High-frequency Intralayer Raman modes.** The high frequency Raman spectra of $ReSe_2$ corresponding to the intra-layer modes are distinctively different from that in other typical TMDCs such as $MoS_2$ and $WSe_2$. Here, we would like to compare the Raman spectra of few-layer $ReSe_2$ with those of few-layer $MoS_2$. First, atomically thin $ReSe_2$ layers have more than ten measured Raman peaks while $MoS_2$ flakes only have two high-frequency Raman active peaks,[45] which are the $E_{2g}^1$ mode and $A_{1g}$ mode. In addition, the lattice vibrations in each $ReSe_2$ Raman mode are much more complex than those in $MoS_2$ layers. It can be easily identified that the $E_{2g}^1$ mode of $MoS_2$ is the in-plane opposite vibrations of two S atoms and one Mo atom, while the $A_{1g}$ mode is the opposite vibration of two S atoms in the vertical direction. For $ReSe_2$, however, it is hard to identify such a pure vibration mode due to the rather complicated crystal lattice and we name the Raman modes based on the dominant direction of phonon vibrations, for example, the $E_g$-like and $A_g$-like modes at 125 $cm^{-1}$, 160 $cm^{-1}$ and 176 $cm^{-1}$ as shown in Figure 3c.

$ReSe_2$ from monolayer to bulk has a $C_i$ symmetry, and the corresponding irreducible representation is $\Gamma = A' + A''$, in which $A'$ is Raman active. Therefore, all the Raman modes in $ReSe_2$ are $A'$ mode with the same format of Raman tensor:

$$A' = \begin{bmatrix} a & d & e \\ d & b & f \\ e & f & c \end{bmatrix}.$$

According to this tensor, the angle-resolved peak intensity of the Raman mode in the polarization measurement is calculated. If the polarization angle in the plane of incident laser is assumed as φ, the intensity of the $A'$ mode $I(A') \propto |a\cos\varphi + d\sin\varphi|^2 + |d\cos\varphi + b\sin\varphi|^2$, see supplement information for the detail. The Raman tensor for each peak of monolayer $ReSe_2$ is different, which may lead to different angle-resolved polarization intensity profile between two peaks. As shown in Figure 3(b), our calculation fits the experimental results well.

Moreover, it is also observed that most of the peak positions for mono-layer and few-layer $ReSe_2$ samples red-shift with increasing layer numbers (Figure 4). Considering van der Waals force only, when the layer number increases, the vdW interaction between layers tends to suppress lattice vibrations, making the lattice more "stiff". Thus, the vibration energy of each vibration mode could become higher with increasing the layer number, resulting in the blue-shifts of the corresponding Raman peaks.[42] On the other hand, if we consider long-range Coulomb interactions, the increased dielectric tensors with the layer numbers can lead to an increase in Coulomb screenings. This results in a softer Coulomb interaction between atoms, and thus the Raman peak tends to redshift.[46] Here, we would like to compare $ReSe_2$ with $WeSe_2$, whose atomic mass is almost equal to $ReSe_2$. The layer-number dependent behavior of high frequency modes in $WSe_2$ is similar with those in $MoS_2$, $A_{1g}$ and $E_{2g}$ modes are blue-shift and red-shift from monolayer to bulk, respectively.[47] The frequencies of C and LB modes in bilayer $WSe_2$ are 16.5 cm$^{-1}$ and 28.2 cm$^{-1}$ respectively,[38] which are higher than those in bilayer $ReSe_2$. The vdW interaction can be measured by the interlayer force constant $\alpha \propto m\omega^2$,[36] where $m$ is atomic mass of monolayer TMDCs, ω is the Raman frequency of interlayer vibration modes. Since the $m$ is almost the same for each other, the interlayer C and LB coupling of $WSe_2$ is about 27% and 49% stronger than that of $ReSe_2$, respectively. This indicates that the interlayer vdW interaction of $ReSe_2$ is much weaker than that of $WSe_2$. Moreover, the long rang Coulomb interaction is estimated by calculating the dielectric tensors and Born effective charges with DFT, see supplement information. The long rang Coulomb interaction in $ReSe_2$ is close to that of $WSe_2$. Based on the above discussions, the observed red-shifts of Raman peaks with increasing $ReSe_2$ layer numbers indicate that interlayer van der Waals interaction is not strong enough to dominate the layer dependence of phonon behavior. The relative weak interlayer coupling of few-layer $ReSe_2$ has also been evidenced by its layer-number dependent PL emission, which will be discussed later.

For both ultralow-frequency and high-frequency in-plane Raman modes, the peak intensities vary periodically with the polarization directions of the incident laser beam, which are a direct evidence of the anisotropy crystal. Once combining high frequency and low frequency Raman measurements, both crystal orientation and layer numbers of $ReSe_2$ flakes can be clearly identified.

**Photoluminescence.** For layered materials, the quantum confinement in the vertical direction can be enhanced once the layer number reduces and the interlayer interaction disappears in monolayer samples. Hence, many TMDC crystals have indirect energy bandgap in their bulk and multi-layer forms, but possess direct bandgap in their monolayer form.[45, 48, 49] The photoluminescence measurements of mono- and few-layer $ReSe_2$ samples revealed that the optical bandgap decreases with increasing layer numbers (Figure 5b), which is similar to other 2D semiconductors, such as $MoS_2$,[50] black phosphorus,[51] and $ReS_2$.[24] Similar to $ReS_2$, the layer-number dependence of the bandgap of few-layer $ReSe_2$ is much weaker than that in other 2D semiconductors. Many TMDCs such as $MoS_2$ and $WSe_2$ have strong PL emission in monolayer flakes with direct bandgap, however, the PL peak intensity of few-layer $ReSe_2$ increases monotonically when increasing the layer number (Figure 5b). The layer-number dependence of the PL peak intensity is much weaker compared to other TMDCs such as $MoS_2$ and $WSe_2$, indicating a relatively weak layer-number dependence of quantum confinement and interlayer interactions. In this case, when adding layers, PL intensity increases due to the increased quantity of materials.

In summary, we report the first experimental observation of the anisotropic interlayer C and LB vibration modes and the anisotropic layer-number dependent photoluminescence in mono- and few-layer $ReSe_2$. A systematic study of the angle-resolved polarization Raman measurements on intra-layer Raman modes of mono- and bilayer $ReSe_2$, as well as their layer-number dependence, is also carried out for the first time. The ultralow frequency Raman modes can be used for the identification of layer number and crystal orientation in $ReSe_2$. The optical bandgaps of $ReSe_2$ are revealed to vary from 1.32 eV in monolayer to 1.26 eV in the bulk, which agree well with the first-principle calculations. Unveiling the interlayer Raman modes and the layer-number dependent optical bandgap in $ReSe_2$ advances the understanding of this 2D TMDC material with unusually complex lattice structure while opening the door to potential applications in near-infrared polarized optoelectronics devices. Finally, the experimental demonstration of anisotropic phonon and optical properties also provides direct experimental evidence of the highly anisotropic nature of mono- and few-layer $ReSe_2$, which may be utilized to construct advanced semiconductor devices. Furthermore, besides tuning $ReSe_2$ bandgap via modifying layer numbers, strain engineering[29] and electrostatic gating[52] might also be utilized to control the energy bandgap. Since the indirect bandgap of $ReSe_2$ is just slightly smaller than the direct inter-band transition energy as shown in the electronic band sturcture, it might be possible to tune $ReSe_2$ into a direct bandgap material with bandgap engineering, which is desirable for various applications like light emission. The tunable bandgap in the near-infrared (NIR) spectrum range can make $ReSe_2$ an interesting material for exploring novel device applications in NIR optoelectonics.

## Methods

**Sample preparation and imaging:** $ReSe_2$ flakes were exfoliated from bulk $ReSe_2$ crystals using standard micromechanical exfoliation method onto high resistivity silicon substrates with 285 nm $SiO_2$. An optical microscope was used to locate few-layer samples via optical contrast. Then the layer numbers were carefully identified by a Dimension Edge AFM.

**High frequency interlayer Raman measurements:** High frequency Raman spectroscopy was measured by a Renishaw InVia spectrometer with 532 nm green laser. In typical measurement, 0.2 mW laser power was applied for 60 seconds through an X100 objective. A half-wavelength plate was used to tune the polarization of the incident laser beam. All the Raman measurements were conducted in dark environment at room temperature.

**Ultralow frequency intralayer Raman measurements: The** Raman spectra at the ultralow frequency was measured in back-scattering at room temperature with a Jobin-Yvon HR800 Raman system, equipped with a liquid-nitrogen-cooled CCD, a X100 objective lens (numerical aperture is 0.90) and 1800 lines/mm gratings. The excitation wavelength was 530 nm from a $Kr^+$ laser. Plasma lines are removed from the laser signals, using BragGrate band-pass filters. Measurements down to 5 $cm^{-1}$ for each excitation are enabled by three BragGrate notch filters with optical density 3 and with FWHM=5-10 $cm^{-1}$. Both BragGrate band-pass and notch filters are produced by OptiGrate Corp.

**Photoluminescence measurements:** Photoluminescence was measured with a Jobin-Yvon HR800 Raman system, equipped with a liquid-nitrogen-cooled InGaAs detector, an X50 objective lens (numerical aperture is 0.55) and 150 lines/mm gratings. The excitation wavelength was 633 nm from a He-Ne laser. A ST500 system was employed to cool the samples to 80K.

**First-principle calculations:** For the calculation of $ReSe_2$ bandstructure, various methods were used and consistent results were obtained. Ultrasoft pseudo-potential was applied, which has been proven to work well for 2D materials. All the energy bandstructure diagrams plotted in this paper were generated by Perdew-Burke-Ernzerhof exchange-correlation function under general gradient approximations, as implanted in the framework of density functional theory (DFT) as implemented in the Vienna ab initio simulation package (VASP). K-space path were selected to go through most of the high symmetric points in reciprocal space, as shown in Fig 1c. Non-collinear spin polarization was considered for SOC related calculations. We found that the neglect of SOC or spin polarization always results in a higher bandgap, though GGA itself typically underestimates bandgap for 2D materials. The projector augmented wave (PAW) method and a plane-wave basis set with an energy cutoff of 500 eV were used in the calculations. For geometry optimizations the Brillouin-zone integration was performed using a regular $5 \times 5 \times 5$ and $5 \times 5 \times 1$ $k$ mesh within the Monkhorst-Pack scheme for bulk and monolayer samples, respectively. The $k$-points have been doubled for the electronic density of states calculations. The convergence criterion of the self-consistent field calculations was set to $10^{-5}$ eV for the total energy. The lattice constants were optimized until the atomic forces are less than 0.01 eV/Å. For monolayer samples, a large vacuum spacing (at least 12 Å) was introduced.

Lattice vibrations for mono-, bi-, and tri-layer $ReSe_2$ were also calculated with GGA-PBE scheme with the k-path shown in Fig. 1b. K-point sampling was 8×8×1 for monolayer sample, and 4×4×1 for bi- and tri-layer ones. For all the calculations carried out in this paper, crystal structures were geometry optimized before any further calculations were conducted.

## Figure Captions

**Figure 1.** (a) The unit cell of $ReSe_2$ crystal. **a**, **b**, and **c** are the three lattice vector. Re atoms and Se atoms are colored with blue and yellow, respectively. (b) the top view of $ReSe_2$ crystal. The clustering of Re atoms forms the Re chains along lattice vector **a** direction, as shown in the red dotted line box. (c) The calculated bandstructure of $ReSe_2$ monolayer. K path are shown in the Brillouin zone above. The arrow indicates the transition from the top of the valence band to the bottom of the conduction band. Fermi level has been set to zero. (d) Left side: the optical and AFM image of monolayer and few-layer $ReSe_2$. Right panel: the AFM height data measured along the pink and blue dotted lines shown on the left.

**Figure 2.** (a) Interlayer Raman modes in monolayer and few-layer $ReSe_2$. C and LB denotes shear and breathing modes. Note that the two peaks around 5 $cm^{-1}$ marked by stars are not Raman signals. They are due to Brillouin scatterings of silicon. (b) Schematic of the three low frequency phonon vibration modes of bilayer $ReSe_2$ obtained from first-principle calculations. From left to right, they correspond to the two shear modes located at 12.1 $cm^{-1}$, 13.2 $cm^{-1}$ and one breathing mode located at 24.8 $cm^{-1}$. (c) Polarization-resolved Raman peak intensity of bilayer (upper panel) and trilayer (bottom panel) $ReSe_2$. The intensity of each peaks have been normalized with respect to the intensity of the Si peak. The dots in both figures are from experiment, while the lines in both figures are theoretical fits from the Raman tensor analysis.

**Figure 3.** (a) Raman spectra of monolayer $ReSe_2$ in the 100 $cm^{-1}$ to 300 $cm^{-1}$ spectral range under the excitation of 532 nm green laser beam at different polarization direction incident normal to the sample. (b) The dependence of Raman intensity on the polarization of the incident laser beam for three typical peaks located at 125 $cm^{-1}$ (red dots), 160 $cm^{-1}$ (blue triangles) and 176 $cm^{-1}$ (green diamonds). The symbols are experimental data while the three curves in corresponding colors are fits from the Raman tensor analysis. (c) The lattice vibration modes of the three peaks shown in (b) obtained from first-principle calculations.

**Figure 4.** Layer-dependent Raman spectra of $ReSe_2$ in the 100 $cm^{-1}$ to 300 $cm^{-1}$ range. The monolayer, bi-layer and 4-layer flakes with the same crystal orientation were measured with the same polarization of the incident laser beam.

**Figure 5.** (a) PL spectrum of $ReSe_2$ in different layers; (b) the layer dependent PL

intensity and bandgap of few-layer $ReSe_2$. (c) Polarization-dependent PL intensity of six layer and ten layer $ReSe_2$.

**Figure 1. Crystal Lattice and Electronic Band Structure of $ReSe_2$**

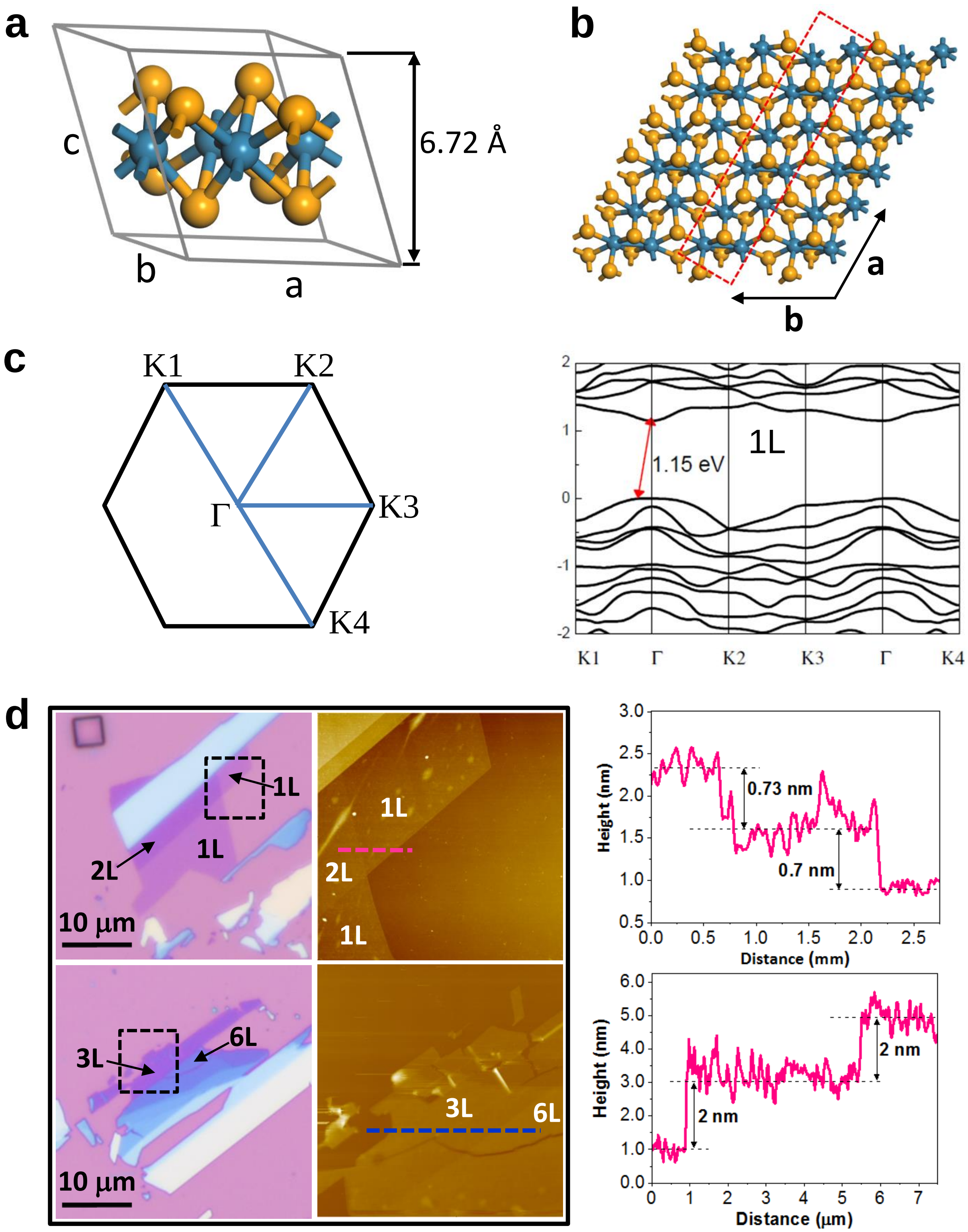

## Figure 2. Interlayer Raman Vibration Modes in $ReSe_2$

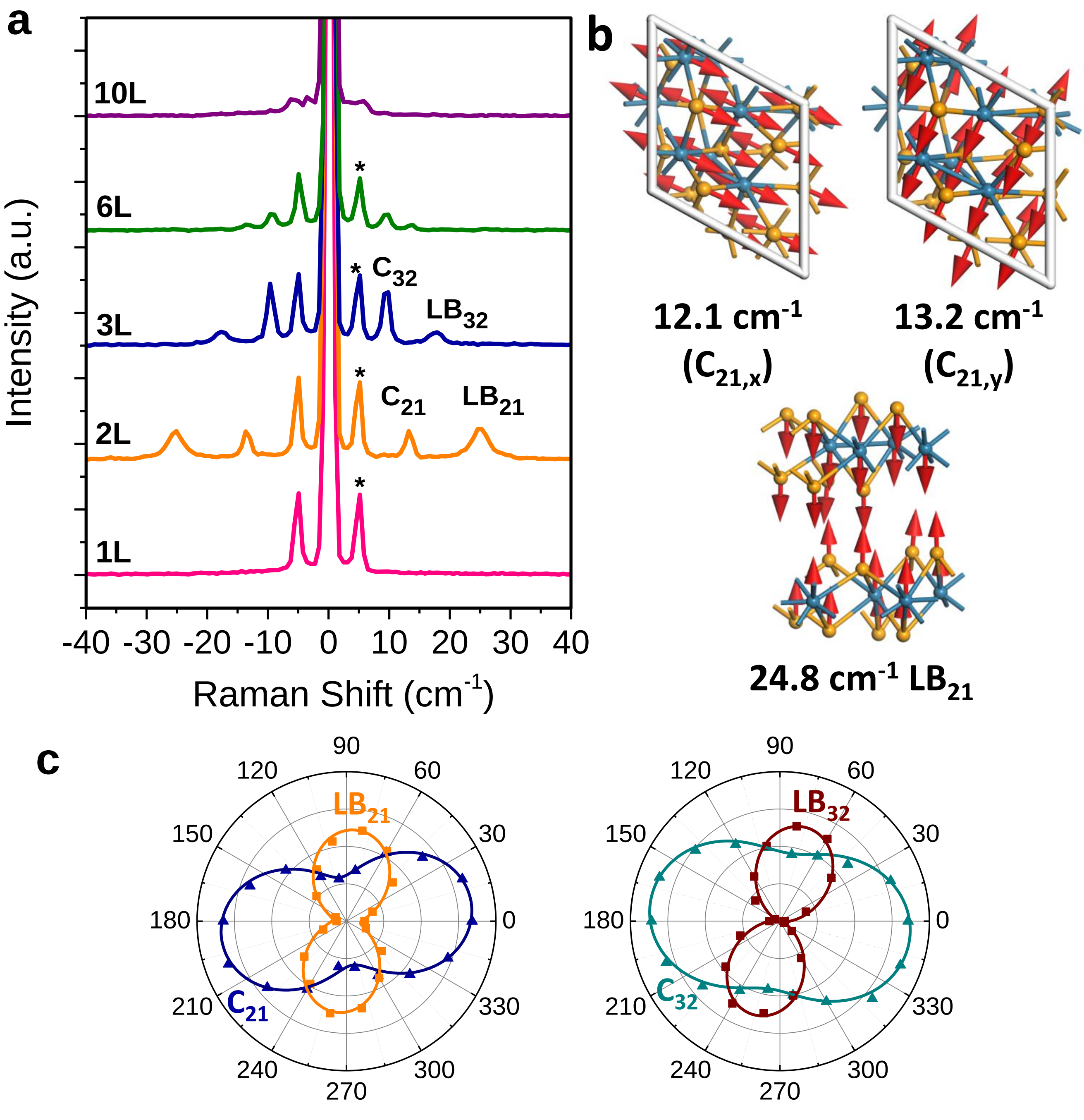

**Figure 3. Intra-layer Raman Vibration Modes in $ReSe_2$**

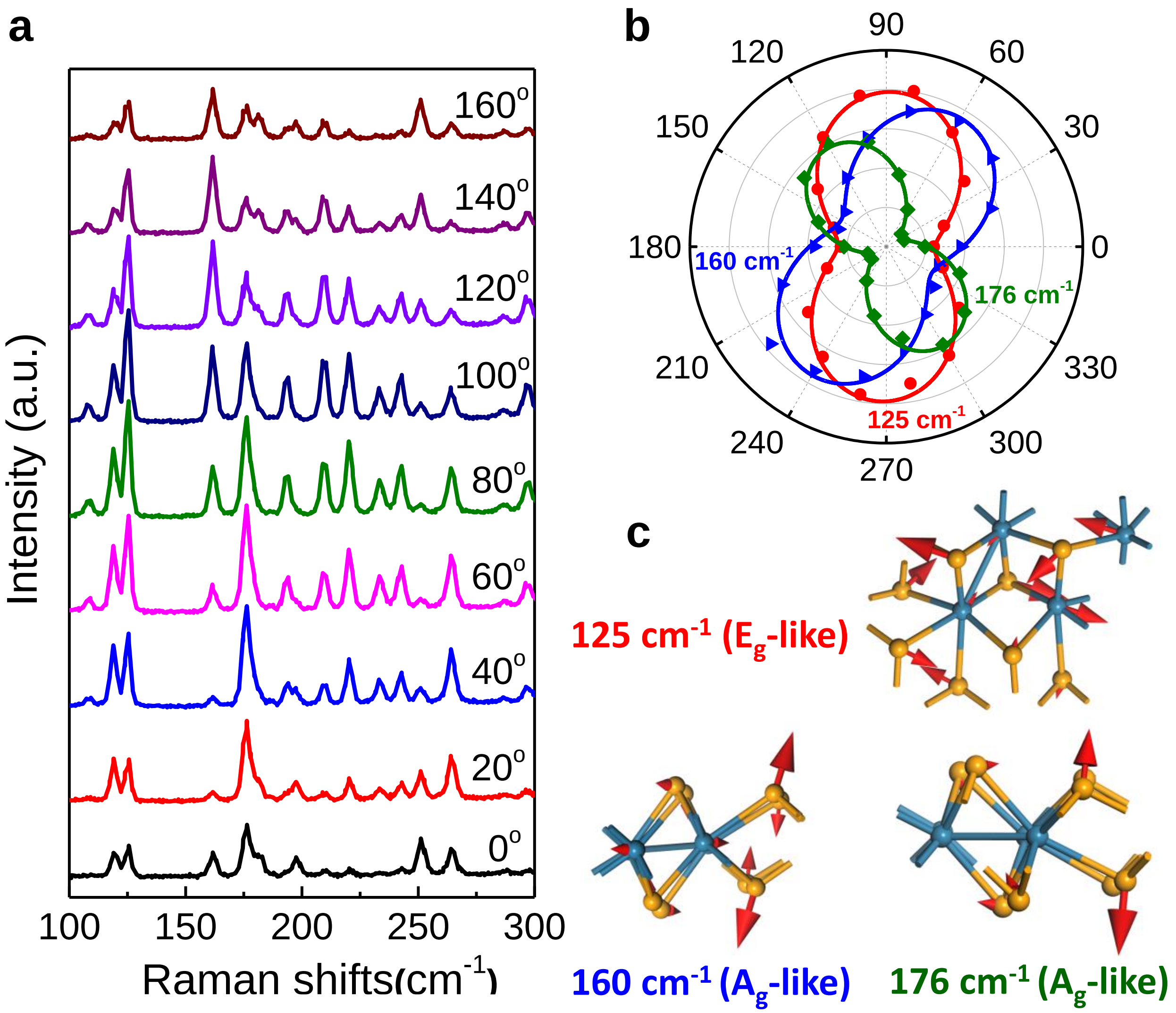

**Figure 4. Layer Dependence of Intra-layer Raman Vibration Modes in $ReSe_2$**

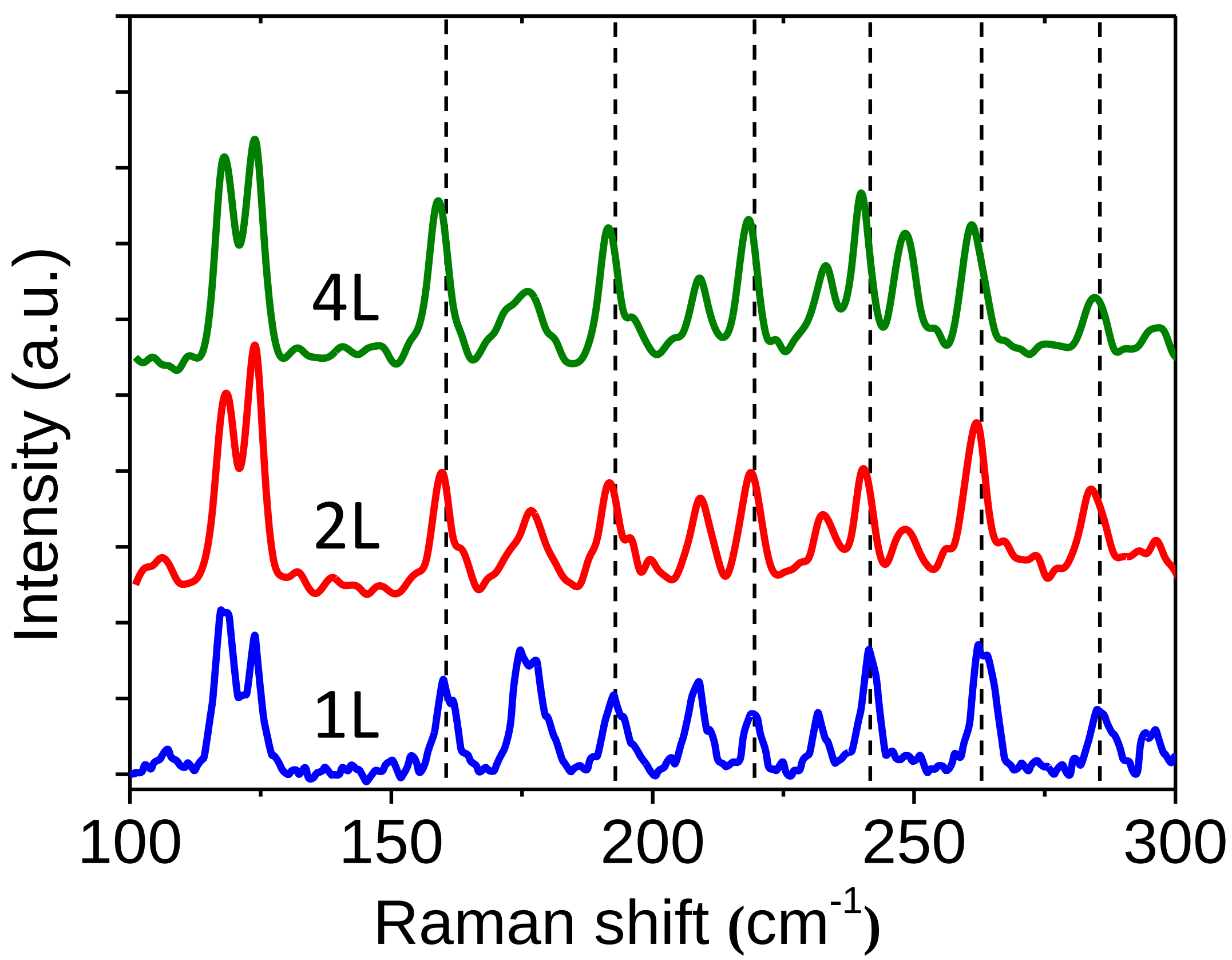

**Figure 5. Photoluminescence Spectrum of $ReSe_2$**

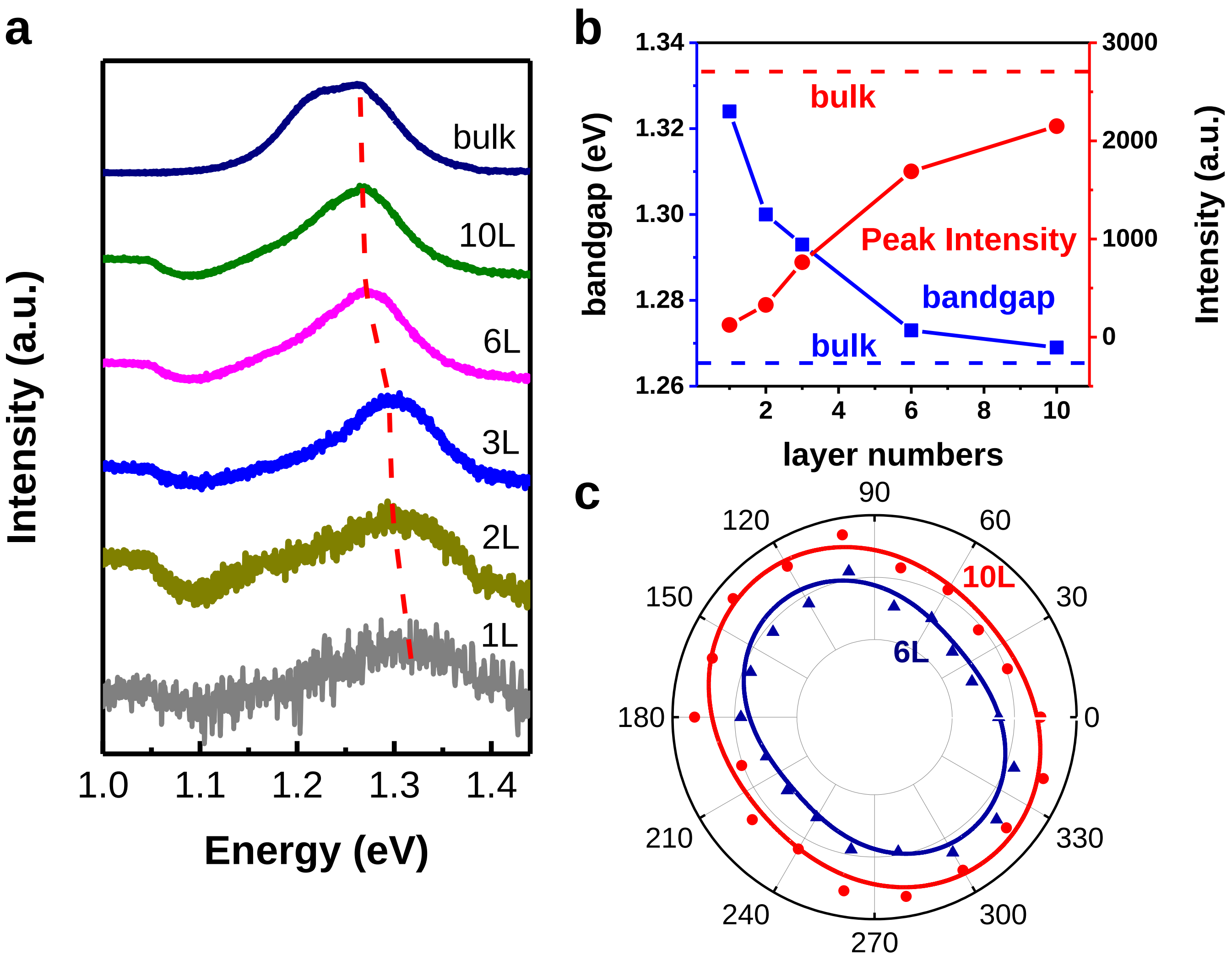

Supplementary Information for

# Interlayer Interactions in Anisotropic Atomically-thin Rhenium Diselenide

Huan Zhao[1,‡], Jiangbin Wu[2,‡], Hongxia Zhong[3,5], Qiushi Guo[4], Xiaomu Wang[4], Fengnian Xia[4], Li Yang[3], Ping-Heng Tan[2,*], Han Wang[1,*]

[1]*Ming Hsieh Department of Electrical Engineering, University of Southern California, Los Angeles, CA 90089, USA*
[2]*State Key Laboratory of Superlattices and Microstructures, Institute of Semiconductors, Chinese Academy of Sciences, Beijing 100083, China*
[3]*Department of Physics, Washington University in St Louis, St Louis, MO 63130, USA*
[4]*Department of Electrical Engineering, Yale University, New Haven, CT 06511, USA*
[5]*State Key Laboratory for Mesoscopic Physics and Department of Physics, Peking University, Beijing 100871, China.*

## 1. Additional discussions about the DFT calculations.

We carried out first principle calculations of the bandstructures for mono-layer and bulk $ReSe_2$ with and without spin orbit coupling considered, keeping the other parameters the same. The calculated bandgap tends to be smaller with SOC effects included for monolayer to bulk $ReSe_2$.

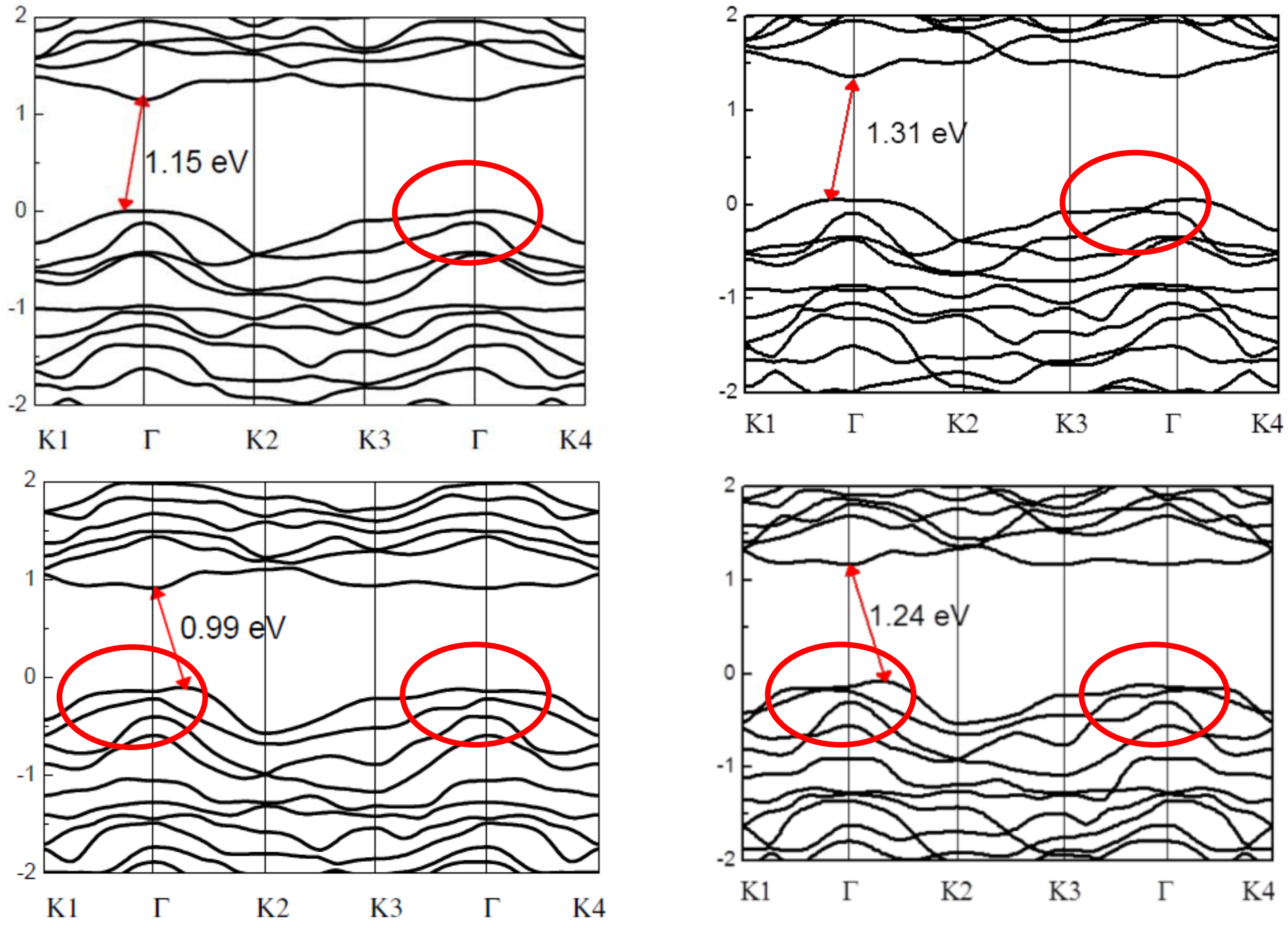

Figure S1. The calculated bandstructures of monolayer and bulk $ReSe_2$. Left top: monolayer with SOC; right top: monolayer without SOC; left bottom: bulk with SOC; right bottom: bulk without SOC. Band spilts are marked by red circles.

**2. The polarization dependent Raman peak intensities of eight typical peaks.**

We observed more than ten peaks in the high frequency (>50 cm$^{-1}$) range of $ReSe_2$ Raman spectrum, all with polarization dependence in the peak intensities. Here, we show eight of these peaks that have the stronger intensities. Different Raman modes have their maximal peak intensity at different polarization of the incident laser beam, which should be attributed to their different dominant phonon vibration directions.

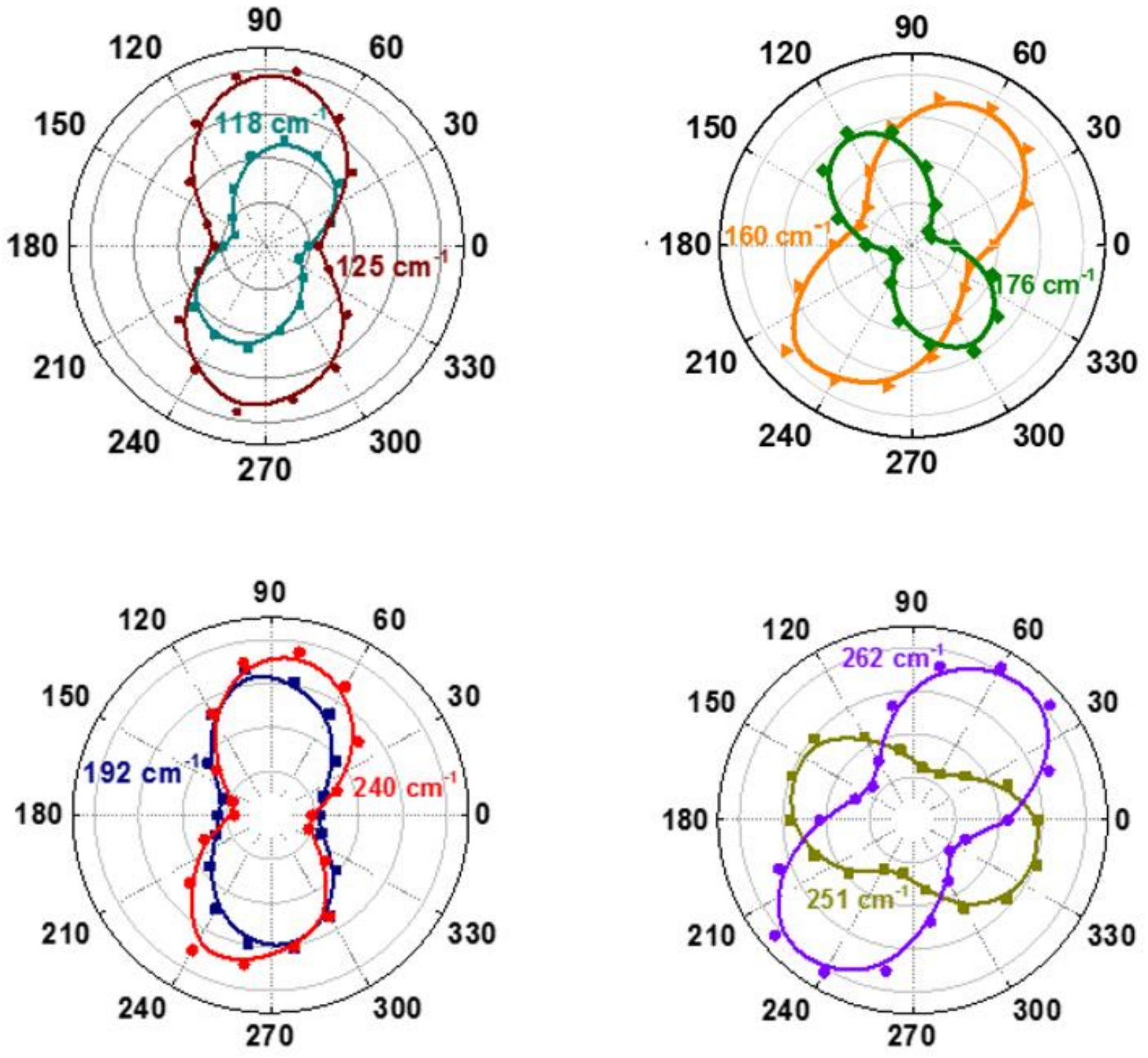

Figure S2. The polarization dependence of Raman peak intensities for eight typical peaks in the 50 cm$^{-1}$ to 300 cm$^{-1}$ range of monolayer $ReSe_2$.

**3. The DFT calculation of low-frequency Raman peaks.**

We have performed DFT calculation of the low-frequency interlayer Raman modes for bi-layer and tri-layer $ReSe_2$. Table 1 is the calculated peak positions and corresponding modes, $C_x$ and $C_y$ mean shear modes along x and y direction, respectively. LB means layer breathing modes. Our calculations can also identify the atom displacements for each mode, as shown in Fig. S3. Some of these modes are inactive in experimental measurements and some are degenerated. The calculated Raman shifts match the experimental data well.

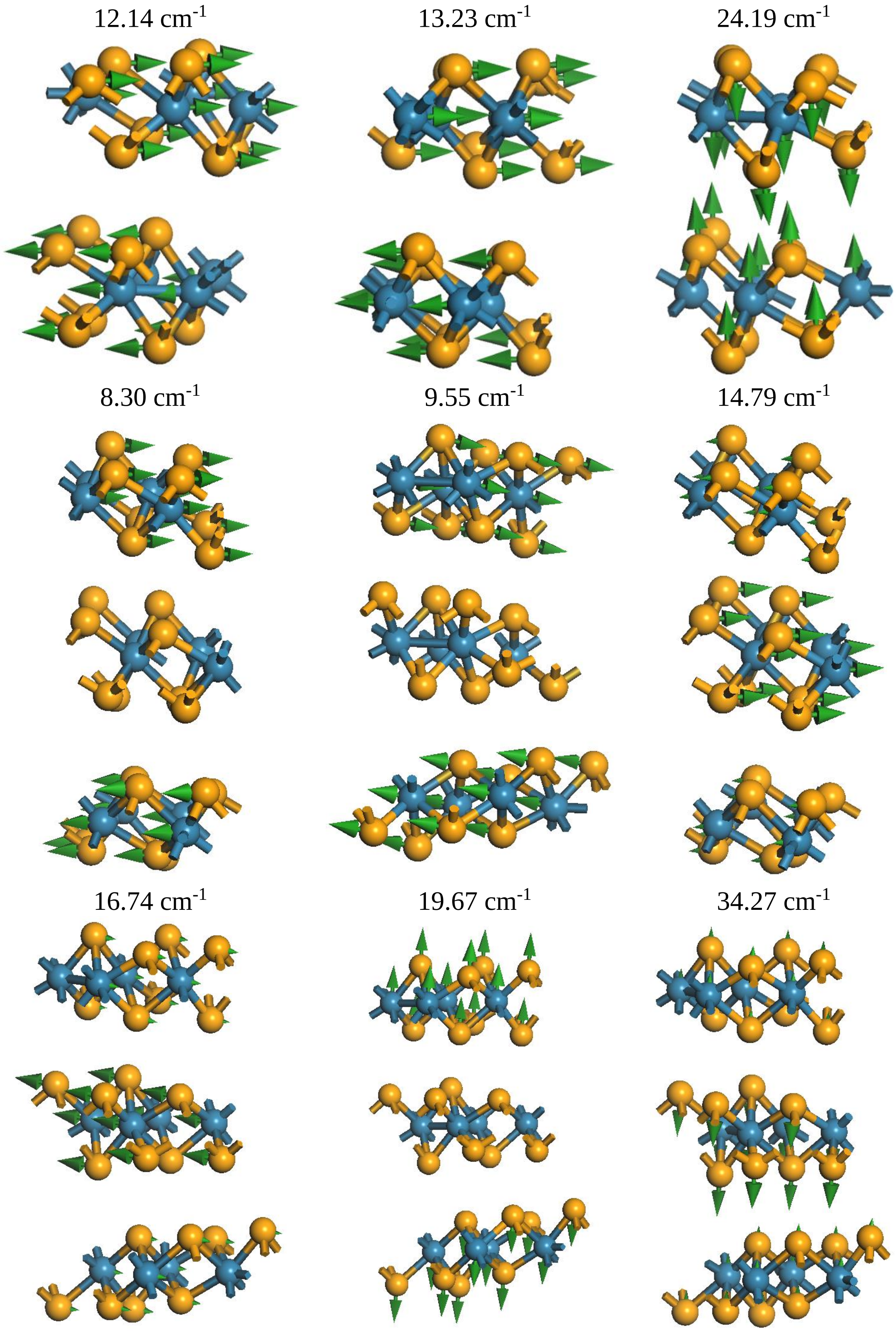

Figure S3：Top: three interlayer vibration modes of bi-layer $ReSe_2$; Middle and bottom: six interlayer vibration modes of trilayer $ReSe_2$.

Table S1: Calculated low frequency vibration modes

| | | | | | | | |
|---|---|---|---|---|---|---|---|
| 2L | Frequency ($cm^{-1}$) | 12.14 | 13.23 | 24.19 | | | |
| | Measured Raman shifts ($cm^{-1}$) | 13.4 (Indiscernible) | | 25 | | | |
| | Mode type | $C_x$ | $C_y$ | LB | | | |
| 3L | Frequency ($cm^{-1}$) | 8.30 | 9.55 | 14.79 | 16.74 | 19.67 | 34.27 |
| | Measured Raman shifts ($cm^{-1}$) | 9.5 (Indiscernible) | | -- | -- | 17.7 | -- |
| | Mode type | $C_x$ | $C_y$ | $C_x$ | $C_y$ | LB | LB |

## 4. The calculated high frequency Raman modes in monolayer $ReSe_2$.

Table S2 is a list of phonon modes of monolayer $ReSe_2$ we obtained by calculations, as well as the corresponding Raman peak positions we observed. The number of total calculated phonon modes is 33, here we only present the ones which we observed in our measurement.

Table S2: a list of calculated phonon modes and observed Raman peak positions of monolayer $ReSe_2$, unit: $cm^{-1}$.

| **Calculated** | **Measured** | | **Calculated** | **Measured** |
|---|---|---|---|---|
| 106.66 | 108 | | 208.61 | 209 |
| 114.63 | 113 | | 218.95 | 219 |
| 117.44 | 118 | | 241.16 | 242 |
| 122.26 | 125 | | 251.77 | 251 |
| 160.95 | 160 | | 261.43 | 262 |
| 175.49 | 176 | | 285.45 | 285 |
| 179.74 | 180 | | 299.23 | 296 |
| 191.18 | 192 | | | |

## 5. Layer-number dependent PL measurement of $ReSe_2$ at 80 K

PL measurement of 3L-, 6L- and bulk $ReSe_2$ were also carried out at 80 K (Fig. S4a). We found that all the peak positions blue-shifted compared to the ones at room temperature for each same sample. We found that the PL peak of all three samples splits into two adjacent peaks with energy difference around 0.02 eV. Similar to the case at room temperature, all the two peaks redshift slightly as the layer number increases. To reveal the PL peak split of $ReSe_2$, we calculated the energy band structure of bulk $ReSe_2$ using

DFT with spin orbit effect considered (Fig. S4b). A calculated band structure of bulk $ReSe_2$ without considering spin orbit effect is demonstrated in Fig. S1 for comparison. Since we use a GGA method, the calculated bandgap tend to be slightly underestimated compared to the measured values.

At low temperature, the PL intensity is weaker than that at room temperature for multi-layer samples, which is also found in some other indirect bandgap TMDCs, such as few-layer $MoSe_2$.[1] In addition, like other TMDCs,[2, 3] the bandgap of all samples increases at lower temperature. This bandgap decreasing at high temperature is universal in semiconductors,[4] due to the increased electron-photon interaction as well as the temperature-dependent lattice length. Taking the bandgap of bulk $ReSe_2$ as an example, the bandgap increased by around 0.1 eV when temperature decreased from 300 K to 80 K. Hence, it would be possible to tune the optical properties of $ReSe_2$ through thermal engineering.

At low temperature, the PL peaks of $ReSe_2$ flakes split into two adjacent peaks with an energy difference of 0.02 eV. The low-energy peak located at around 1.37 eV always has higher intensity. For example, in the 6L $ReSe_2$ flake, the peak at 1.37 eV has a full width at half maximum (FWHM) of around 0.03 eV, and the peak at 1.39 eV has an FWHM around 0.02 eV. At room temperature, however, the 6L $ReSe_2$ flake sample has an FWHM of 0.11 eV, which is much larger than that in the two separate peaks at low temperature. The PL peaks of all the other samples also exhibit a significantly broadened FWHM at higher temperature. Hence, the PL peak at room temperature is most likely the combination of multiple peaks. From Fig. S4b and Fig 1c of the main text, the valence band of both monolayer and bulk $ReSe_2$ near Γ point is relatively flat. The indirect inter-band transitions start from the top of valence band, which is close to Γ point and is 0.01-0.03 eV higher in energy than the highest valence band energy at Γ point. At room temperature, thermal fluctuation energy is in the scale of kT, where k is Boltzmann constant and T is room temperature (~300 K). The estimated thermal fluctuation energy is 0.03 eV. This energy is large enough to modify the energy of some of the electrons at Γ point to the extent of 0.01-0.03 eV, and thus may trigger the direct inter-band transitions. Hence, the transitions excited by the laser excitation may happen in multiple positions of the valence band near Γ point, resulting in a broadened PL spectrum.

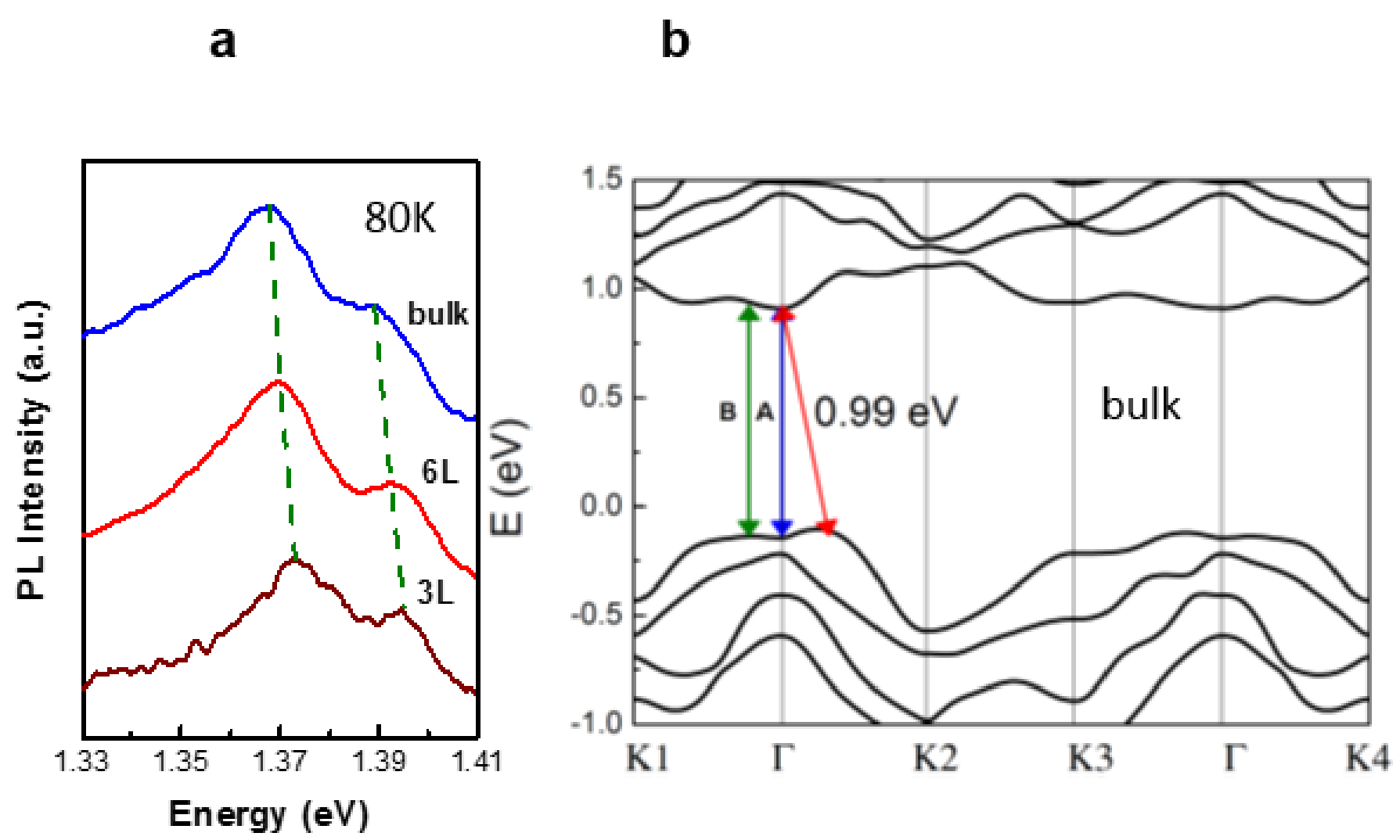

Figure S4 (a) PL spectrum of $ReSe_2$ samples at the temperature of 80K. The intensities of each peaks are rescaled to fit the figure. (b) The calculated bandstructure of bulk $ReSe_2$. The bottom of the conduction band is at Γ point, whereas the top of the valence band is near the Γ point. Those local maxima away from the Γ point can contribute to van Hove Singularities. The energy of the transition B is about 20 meV higher than that of the transition A at the Γ point.

The energy difference between the two split PL peaks at 80 K is 0.02 eV. Since the thermal fluctuation energy at room temperature is higher than 0.02 eV, it is understandable that these two peaks cannot be distinguished at room temperature. It is well-known that $MoS_2$ PL has two peaks with energy difference of 0.15 eV due to spin orbit coupling (SOC) induced split of valence band. From our calculation, the SOC induced band splitting does not occur at the transition position, so the split of PL peak cannot be attributed to SOC effect. In addition, the SOC induced energy split is over 0.1 eV according to the calculation, which is much larger than the split energy we observed. Impurities and defects can also lead to PL peak split, but they are not likely to contribute to the split we observed since we consistently observe two peaks with stable peak positions in different samples with different layer number. Furthermore, as we can observe similar peak split in the bulk sample compared to in trilayer sample, it is unlikely that the split is due to trion or exciton PL peaks in $ReSe_2$ samples. From the band structure (Fig. S4b), bulk $ReSe_2$ is an indirect bandgap semiconductor. The bottom of the conduction band is at Γ point, whereas the top of the valence band is near the Γ point. It is the most likely that those local maxima away from the Γ point contribute to van Hove

Singularities, giving rise to the side peak and corresponding splitting. As marked in Fig. S4b, the energy of the transition B is about 20 meV higher than that of the transition A at the Γ point. This is consistent with experimental observation in Fig. S4a. Meanwhile, consider the small energy difference (around 20 meV), phonons may play an important role for the observed peak splitting.

### 6. Symmetry analysis and Raman tensor of $ReSe_2$

The ReSe2 crystal has a Ci symmetry, hence the corresponding irreducible representation can be denoted by Γ=A'+A'', where A' mode is Raman active, and A'' is infrared active. In our measurements of ReSe2, all of the Raman peaks are A' mode. The Raman tensor of the A' mode can be written as:

$$\mathrm{A}' = \begin{bmatrix} a & d & e \\ d & b & f \\ e & f & c \end{bmatrix}.$$

Raman intensity is proportional to |ei·R·es |2, where ei and es are the unit vectors describing the polarizations of the incident and scattered lights, R is Raman tensor. In this work, the polarization of the incident light was tuned with an angle ϕ by a half wavelength plate (es =[cosϕ sinϕ 0]), while the polarization of the scattered light are free. Thus, the intensity of Raman peak in ReSe2 is:

$$\mathrm{I} \propto \left| \begin{bmatrix} cos\varphi & sin\varphi & 0 \end{bmatrix} \begin{bmatrix} a & d & e \\ d & b & f \\ e & f & c \end{bmatrix} \right|^2 = |a \cos \varphi + d \sin \varphi|^2 + |d \cos \varphi + b \sin \varphi|^2$$

Therefore, the polarization-resolved intensity profile of the Raman peaks depend on the relative values of $a, b$ and d. These values in each mode are determined by how the phonons vibrate, offering each peak a different polarization-resolved intensity profile. For both interlayer Raman modes and intralayer Raman modes, we were able to get the polarization dependence of peak intensity via fitting the values of $a, b$ and d. Our calculations match the experiments well (Figure 2c and Figure 3b).

### 7. Analysis of the long rang Coulomb interaction

According to a previous work,[5] the long rang Coulomb interaction can be estimated as:

$$C^{lr}_{M,x\ M,x} = -2 \frac{(Z^*_{M,xx})^2}{\sqrt{\epsilon_{xx}\epsilon_{zz}}d^3},$$

where $Z^*_{M,xx}$ is the Born effective charges tensor of metal atom $M$, $\epsilon_{xx}$ is the dielectric tensors, $d$ is the distance of two atoms. The ratio of long rang Coulomb interaction between monolayer and bulk can be used to measure the strength of Coulomb screenings:

$$\frac{C^{lr}_{M,x\,M,x}(mono)}{C^{lr}_{M,x\,M,x}(bulk)} = (\frac{Z^*_{M,xx,mono}}{Z^*_{M,xx,bulk}})^2 \sqrt{\frac{\epsilon_{xx,bulk}\epsilon_{zz,bulk}}{\epsilon_{xx,mono}\epsilon_{zz,mono}}}.$$

Form the DFT calculation, the Born effective charges $Z^*_{M,xx}$ are almost equal for both systems in $WSe_2$, and the average Born effective charges of monolayer $ReSe_2$ is about 8% more than that of bulk. The dielectric tensor of bulk $WSe_2$ is $\epsilon_{xx,bulk}/\epsilon_{zz,bulk}$=14.68/8.15, and monolayer is $\epsilon_{xx,bulk}/\epsilon_{zz,bulk}$=4.60/1.29. The dielectric tensor of bulk $ReSe_2$ is $\epsilon_{xx,bulk}/\epsilon_{zz,bulk}$=15.70/7.82, and monolayer is $\epsilon_{xx,bulk}/\epsilon_{zz,bulk}$=6.12/1.47. Therefore, the $C^{lr}_{M,x\,M,x}(mono)/\,C^{lr}_{M,x\,M,x}(bulk)$ is equal to 4.49 and 4.29 for $WSe_2$ and $ReSe_2$ respectively. This indicates that the strength of Coulomb screenings in these two materials is quite closed.